\documentclass[12pt]{article}
\usepackage{sproclmo}
\usepackage{epsfig}
 \newcommand \be {\begin{equation}}
\newcommand \ee {\end{equation}}
 \newcommand \bea {\begin{eqnarray}}
\newcommand \eea {\end{eqnarray}}
\newcommand \nn \nonumber
\def \(({\left(}
\def \)){\right)}

\title{  Possible Observation of Nuclear Reactor Neutrinos  Near the  Oscillation  Absolute Minimum.
 \centerline{\it  24 April    2003} }
\begin{document}
  \bibliographystyle{unsrt}
\author{C. BOUCHIAT }

\address{Laboratoire de Physique Th\'eorique de l'Ecole Normale Sup\'erieure
\footnote {UMR 8549: Unit\'e Mixte du Centre National de la Recherche Scientifique
 et de l'\'Ecole Normale Sup\'erieure}
24, rue Lhomond, F-75231 Paris Cedex 05, France }

  \maketitle\abstracts{ We begin by a r\'esum\'e of the basic  $3 \,\nu $  neutrino  oscillation formalism.
 We review briefly  our present empirical knowledge of the oscillation parameters and conclude that  the
   $2 \,\nu $ model  is adequate to describe the survival probability $ P( {\nu}_e  \rightarrow {\nu}_e) $.
   Then we proceed to the evaluation  of  $ P( {\nu}_e  \rightarrow {\nu}_e) $  for the antineutrinos
   emitted by the nuclear power plants presently in operation along the Rh\^one valley. We assume that a
   detector has been installed in an existing cavity located under the Mont Ventoux at a depth
   equivalent to  $1500 \,m  $ of water. We show that such an experiment would provide the opportunity
   to observe neutrinos near the oscillation absolute minimum. We end by a rough estimate of the counting
   rate.}
\section{Introduction}
In this note we would  like to investigate  a possible way to observe the $ {\bar{\nu} }_e  $  oscillations
near   the absolute minimum   of    the survival  probability $   P( {\bar{\nu} }_e  \rightarrow {\bar{\nu} }_e)  $.
 If one uses the $ {\nu}_e $  oscillation parameters deduced  from a global analysis of  the  SNO solar
neutrino events  \cite{sno}  combined  with the reactor antineutrino events
observed with the KamLAND detector \cite{kamland}, then the absolute minimum
of $ P( {\bar{\nu} }_e  \rightarrow {\bar{\nu} }_e)  \simeq  0.2 $ is reached  for  a reactor-detector distance  $ \simeq  70\, km $.
In comparison, the  average distance  $ \simeq  180 \, km $   in the KamLAND experiment leads to a survival probability
  $ P( {\bar{\nu} }_e  \rightarrow {\bar{\nu} }_e)  \simeq  0.6 $.
  \section{Basic Neutrino Oscillation Formalism}
  The neutrino  oscillations  are described \cite{buch,bilen}  by  the      energy eigenstate  $ | {\nu}_{\ell} E \,z  \rangle $
    associated with neutrinos propagating along the z axis and  satisfying   the  boundary condition :
   $   |  \nu_{\ell}\,E E \,z =0 \rangle =   |  \nu_{\ell}\,E \rangle $,   where  $ |  \nu_ {\ell} \,E\rangle $   is the neutrino   state emitted at $z=0$
   by the charged lepton $\ell$ with $ \ell =e,\,\mu,\,\tau $.     The state  $ | {\nu}_{\ell} E \,z  \rangle $   is then  written  as a superposition   of the
   mass matrix  eigenstates    $  |\nu_{i} \rangle \,E  $   relative to  the mass $ m_i $  :
   \be
  | {\nu}_{\ell} E \,z  \rangle = \sum_{1}^{3} | \nu_{i}\,E \rangle \exp(\,i \,p_{iz}  \,z )    U^{*}_{\ell,i}
\ee
where $  p_{iz} \simeq   E- {m_i}^2 /( 2 \, E) $ and $ U  $ the $ 3 \times 3$ mixing matrix, which is usually  written as follows :
  \be
 U=   U_1 \cdot  U_2   \cdot    U_3
  \ee

 \be
 U_{3}= \left( \begin{array} {clcr}
       \cos ({{\theta}_{12}})    &  \sin ({{\theta}_{12}})  &   0 \\
      - \sin ({{\theta}_{12}})   &  \cos ({{\theta}_{12}}) &   0  \\
        0                      &   0                   &   1
      \end {array}    \right)
 \ee

  \be
 U_{1}= \left( \begin{array} {clcr}
       1                      &  0                      &  0    \\
       0                      &  \cos ({{\theta}_{23}})   &  \sin ({{\theta}_{23}})  \\
       0                      & - \sin ({{\theta}_{23}})  &  \cos ({{\theta}_{23}})

      \end {array}    \right)
 \ee

 \be
 U_{2}= \left( \begin{array} {clcr}

        \cos ({{\theta}_{13}})   &  0            &  \sin ({{\theta}_{13}})  \\
        0                      &  1            &  0                     \\
       - \sin ({{\theta}_{13}})  &  0            &  \cos ({{\theta}_{13}})

      \end {array}    \right)
 \ee
 ( Note we have ignored, for the moment, the possibility of   CP violation.)
 In the particular case of two neutrino oscillations,  it is of interest to note
 the correspondence  with the propagation of  photons with energy E in  a birefringent medium
 where the  indices associated  to  the linear polarizations taken    along the two optical axes are given by:
  $ n_i \simeq  1 - {m_i}^2 /( 2 E^2  )\;  (i = 1\,,2)   $.   The  polarization eigenstates  correspond
  to  the mass matrix eigenstates  $    | \nu_{i} \rangle   $.
 It is of convenience to introduce    the oscillation length $L_{osc} $ which is  defined as:
 \be
    L_{osc}( E, \,\Delta \,m^2) = \frac{ 4  \hbar\,c E }{ \Delta \,m^2}
\ee
 Assuming no CP violation, the  antineutrino survival probability at distance $L$  from the detector
  $  P( {\bar{\nu} }_e  \rightarrow {\bar{\nu} }_e,L )=   \langle  {\nu}_{\ell} E \,L  | {\nu}_{\ell} E \,0  \rangle $
  can be written under the form:
  \bea
   P(3\nu| {\bar{\nu} }_e  \rightarrow {\bar{\nu} }_e,E,L) & = & 1 - 4\,\left( {\cos}^2 ({\theta}_{12})\,{\cos}^4 ({\theta}_{13})\,
      {\sin}^2 (\frac{L}{L_{osc}(E,{\Delta }_{12}) }   )\,
      \sin^2 ({\theta}_{12}) \right)  \nn \\
  & & -4\,\left(     {\cos}^2 ({\theta}_{12})\,{\cos}^2 ({\theta}_{13})\,
      {\sin}^2 ( \frac{L}{L_{osc}(E,{\Delta }_{13}) }  )\,
      {\sin}^2 ({\theta}_{13})  \right) \nn  \\
    &  & -4\,\left(   {\cos}^2 ({\theta}_{13})\,
      {\sin}^2 (\frac{L}{L_{osc}(E,{\Delta }_{23})}\,
      {\sin}^2({{\theta}_{12}})\,{\sin}^2 ({{\theta}_{13}}) \right)
      \label{3nuPsurv}
     \eea
      In the above expression   $   {\Delta }_{i_1\,i_2}  $ stands for the mass square difference  $ m_{ i_2}^2- m_{ i_1}^2   $.

      Fogli {\it et al} \cite{fog1} and Bahcall { \it et al } \cite{bah}  have performed     global analysis  which combines
       the  solar neutrino SNO \cite{sno}  experimental results   \cite{sno}
         with  the reactor antineutrino events   observed  in the KamLAND  \cite{kamland} liquid scintillator  detector,
         using essentially the survival probability given in eq. (\ref{3nuPsurv}).
       We quote  here the "summary"  of the current   $ 3 \nu $ situation, as given by Fogli {\it et al} \cite{fog1} :
         \bea
          {\Delta }_{12} &=& ( 7.3\pm 0.8 )\times 10^{-5} \, eV^2 \nn \\
           \sin^2 ({\theta}_{12})&=& 0.315 \pm0.035      \nn \\
           \sin^2 ({\theta}_{13})&  \leq & 0.017
           \label{paretomu}
          \eea
          The quoted errors correspond to one standard  deviation.   It should be said that a similar bound upon
         $  \sin^2 ({\theta}_{13}) $  has been  obtained by combining  the  results from the  Chooz reactor experiment \cite{chooz}
         with  those from the    atmospheric neutrino SK  observations \cite{sk}.

            Fogli {\it et al} \cite{fog2}  have also analysed recently   the atmospheric neutrinos SK \cite{sk}
           $ {\nu}_{\mu} \rightarrow {\nu}_{\tau}$  flavor transition in
          combination with the preliminary results of the  K2K \cite{k2k}   accelerator neutrino experiment .
          They  arrive to the following determination  of  the  relevant oscillation  parameters:
          \be
            {\Delta }_{23}= (2.6\pm 0.4)  \times 10^{-3}\,eV^2  ;\;  \sin^2 ({\theta}_{23})=1.00^{+0.00}_{-0.05}
             \label{parmutotau}
             \ee
            It is easy to verify that   with  the above values  of the oscillation parameters
            the survival probability  $P( {\bar{\nu} }_e  \rightarrow {\bar{\nu} }_e)$
            can be replaced, up to corrections   $  \leq  3 \% $, by the following two neutrino approximation:
               \be
                  P( 2 \nu |\, {\bar{\nu} }_e  \rightarrow {\bar{\nu} }_e,E,L)  =  1 - 4\,\left( {\cos}^2 ({\theta}_{12})\,
      {\sin}^2 (\frac{L}{L_{osc}(E,{\Delta }_{12}) } ) \sin^2 ({\theta}_{12}) \right)
                \label{2nuPsurv}
               \ee
              \section{Survival Probability for Rh\^one Reactor Neutrinos  with a Detector Located under the Mont Ventoux}
\begin{figure}[t]
\centerline{ \epsfxsize=120mm\epsfbox{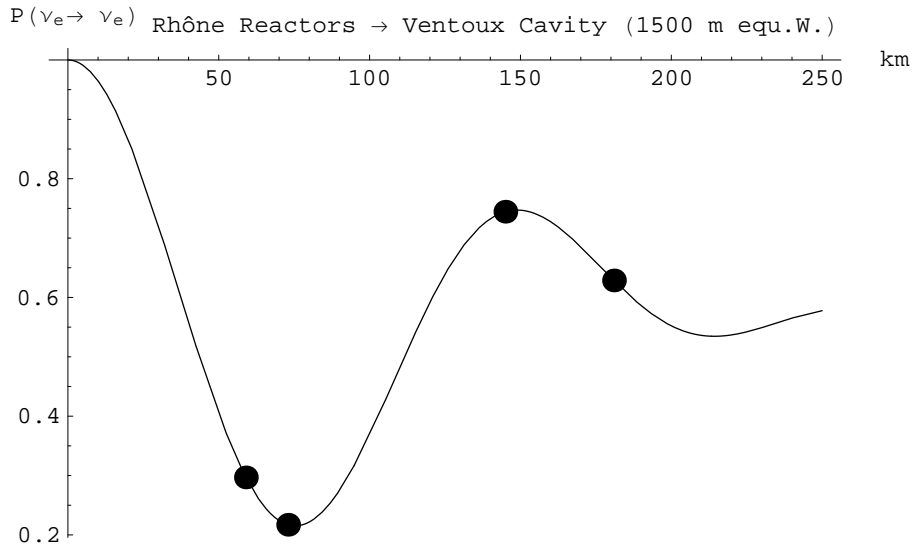} }
\caption{ \footnotesize The curve of  this figure  gives  the average survival probability
of an  antineutrino detected  at the distance $ L $ from the reactor obtained in the $ 2 \, \nu $ approximation,
with the parameter values:  $ {\Delta }_{12} = 7.3\times 10^{-5} \, eV^2 $   and    $  \sin^2 ({\theta}_{12})=0.315 $.
We have assumed that a detector  is to  be  installed in a cylindrical horizontal cavity
( diameter $ = 8 \, m$  with a total length $  = 28 \, m $, including the two hemispherical end caps ),  located under the Mont Ventoux.
 The limestone  vertical depth   is equivalent to  $ \sim  1500 \,m $ of water. The four points on the curve correspond to the four
 nuclear power stations, presently in operation along the Rh\^one river. Their average electric
 power, as given by the EDF company,  is about 2.5  GW ,  close to that of the Chooz site, used  in previous neutrino oscillation experiments.  }
  \label{fig1}
\end{figure}

            We would like now to compute the  probability   $ P_D(E,L) \,dE $    of an antineutrino  to be
            observed at a distance $ L$  from the nuclear reactor and  having  its  energy in the range $ E,E+dE  $.
            The detector is  assumed to be  of  the liquid scintillator type,
            as in the Chooz \cite{chooz} and the KamLAND \cite{kamland} experiments.
              $ P_D(E,L) $ is given as the product  of three   factors. The first factor is the   survival   probability
               $ P( 2 \nu |\, {\bar{\nu} }_e  \rightarrow {\bar{\nu} }_e,E,L) $. The second one  is the energy   spectrum $ S_{\nu} (E)$
                of the antineutrino emitted  by the reactor,    obtained by combining with appropriate weights the spectra associated with the various
                $\beta^{-}$  decays involved in the the fission process. The third term is  the cross section $ {\sigma}_{\nu}(E) $ for the
                inverse  $\beta $  decay reaction: $  {\bar{\nu} }_e +p \rightarrow e^{+} + n    $. The physical quantity measured in the detector is
                the   prompt energy $ E_{pr} $ resulting from the annihilation of the positron with the electrons of the  liquid scintillator:
                   \be
                     E_{pr}=E_{ e^{+} }+m_e\, c^2 = E_{ {\bar{\nu} }_e }- (m_n-m_p)\,c^2 +T_n + m_e\, c^2
                      \ee
                 where $ T_n $ is the  kinetic energy of the recoiling neutron  which can be neglected in our approximate treatment.
                 The antineutrino energy   $  E_{ {\bar{\nu} }_e }  $ is then given by   $ E_{ {\bar{\nu} }_e } \simeq E_{pr} +0.78 MeV $.
                 We have  deduced  the detection probability in the  absence of oscillations:
                 $ P_{D}(E,0) =  P_{pr}(E- 0.78 MeV) \propto S_{\nu} (E) \, {\sigma}_{\nu}(E)$,
                   from the  "expected" prompt energy spectrum  $ P_{pr}(E_{pr} ) $ given  in  the KamLAMD paper \cite{kamland}.
                 We have kept  the same cut  $E_{ pr} \geq  2.26  MeV $, introduced
                 in order   to eliminate the events associated with the "geoantineutrinos". In our computation,  the  log. of the neutrino
                 spectrum,  has been approximated  by
                  a  polynomial  fit: $ \log(  P_{D}(E,0 )) = -2.982+1.141\, E -0.1611 E^2 $ with $ P_{D}(E,0)$
                 normalized to unity in the interval $ 3.38 MeV \leq E  \leq  8 MeV$.
                  The energy-average survival probability is  then  obtained by
                 a  simple numerical quadrature:
                 \be
                  P(  {\bar{\nu} }_e  \rightarrow {\bar{\nu} }_e,L)= \int dE   \, P( 2 \nu |\, {\bar{\nu} }_e  \rightarrow {\bar{\nu} }_e,E,L) \, P_{D}(E,0)
                 \ee
                  The  curve  $  P(  {\bar{\nu} }_e  \rightarrow {\bar{\nu} }_e,L)$       versus $ L$   is plotted in Figure \ref{fig1}.  The  parameters  $ {\Delta }_{12}$
                   $\sin^2 ({\theta}_{13})$  used in the computation  have  the central values  given  in eq. (\ref{paretomu})    and  (\ref {parmutotau}).

                  \begin{tabular}[t]{| l | l | l | }
             \hline
                  & Distance to detector  & Average Electric Power \\
             \hline
             Bugey & 181 \,km &  2.66 \,GW\\
             \hline
             Saint Alban & 145 \,km & 1.92 \, GW \\
             \hline
             Cruas        & 73 \, km   & 2.84 \, GW  \\
             \hline
             Tricastin    & 59 \, km   &  2.59 \, GW \\
             \hline
             \end{tabular}   \\

               Table 1.{ \footnotesize Nuclear power  plants in operation along  the Rh\^one River  with  their distance to
               a detector supposed  to be installed in the Mont  Ventoux cavity and their average electric power.}

             In  each of  two  easily accessible sites,  located  in the  Vaucluse   Department,   there exists  an   identical  network of $3 \,km $
               subterranean galleries,  which were  bored for the French  Ministry of Defense. They are no longer used for military
              purposes. They both contain  the same  cavity  with  concrete inner $ 2.1\, m $  thick  walls. The  inside volume  has the  form of  a cylinder
               with a  $ 8 \,m $ diameter, terminated by the two hemispherical end-caps. The total length of the cavity is $ 28\,m $.
                   The limestone  $ 500\,m $ vertical depth   is equivalent to  $ \sim  1500 \,m $ of water.
               The first site,  accessible from the Rustrel village, is occupied by  the LSBB laboratory ( Le Laboratoire Souterrain Bas Bruit; www.lsbb.univ-avignon.fr )
               The second site is located  near the  Reilhanette village  in the Mont Ventoux  foothills. It  is presently unoccupied. Most of the equipments
                necessary to run a laboratory  have been  removed and the  gallery entrance is closed by a concrete wall. So a non negligible
              initial   investment would be  necessary if one wishes  to install  a neutrino detector in this cavity.

  In the following, we shall assume that an antineutrino detector has  been built in the Mont Ventoux cavity.  The distances $ d_i \; ( 1\leq i \leq 4) $
  from the   Rh\^one valley four power plants  are given in Table 1.  The big points appearing on the curve in Figure 1. correspond
   to the survival probabilities  $  P(  {\bar{\nu} }_e  \rightarrow {\bar{\nu} }_e,d_i )  $. We see clearly that the two   closest  reactors
     ( Tricastin and  Cruas ) will allow the study of  $  P(  {\bar{\nu} }_e  \rightarrow {\bar{\nu} }_e,L) $  near the oscillation absolute
     minimum, a region which has not be explored by the previous reactor experiments. Is is also instructive  to compute the
     survival probability averaged over the four nuclear power plants. We shall assume that the  corresponding    antineutrino flux  is proportional
     to the average electric power  $ \ W_i $ delivered  by each power plant, as given in Table 1.. We get in this way :
     \be
     \langle    P(  {\bar{\nu} }_e  \rightarrow {\bar{\nu} }_e ) \rangle = \sum_1^4 W_i \, d_i^{-2}    P(  {\bar{\nu} }_e  \rightarrow {\bar{\nu} }_e,d_i )/
     (   \sum_1^4 W_i \, d_i^{-2} )=  0.3142
     \label{pnuetonueav}
      \ee
      We see that adding the contributions of the  two far away power plants  leads to an average distance which is still
      in the vicinity of   the absolute minimum.

   \section{Counting Rate Estimate}
We would like to end this note by giving  a rough estimate of  the number of neutrino events per day, assuming that  the liquid scintillator   is contained
in an horizontal cylinder having a $ 6 \, m$ diameter terminated by  two hemispheric end-caps.\nopagebreak
The total length is  taken to be  equal to be $ 22 \,m $. The
volume occupied by the liquid scintillator is then  $V_{Re}= 565 \, m^3 $.  It  is about  one hundred
times larger than the corresponding volume  in  the Chooz experiment  \cite {chooz}   $V_{Ch}= 5.55 \, m^3 $.
( Some experimenters  will probably find that I am  overoptimistic !)
We use here the electric power instead  of the thermal power
so that the average signal given in the Chooz paper \cite{chooz} eq.( 46)  should read    $ X_{Ch} =7.56 $ counts per day per  GW .
We are going to assume the Mont Ventoux detector to have  the same efficiency as the Chooz one; this implies  that the counting rate is proportional
 to the  liquid scintillator volume.   For a given reactor, the antineutrino flux, $ {\Phi}_{ {\bar{\nu} }_e }$ is given, up to a factor  assumed to be the same
for all the French nuclear reactors, by the ratio of the  electric power $ W_i $ to the  square of the distance $d_i^2$. The total flux is given
by  $ {\Phi}_{ { \bar{\nu} }_e ,\,Re  }  = \sum_1^4 \,  W_i   d_i^{-2} =0.0145 \, GW\, km^{-2} $.    The antineutrino flux  leading  to
the counting  rate   $ X_{Ch}  $ is  equal to one if one uses the same units and
 remembers that the distance form the reactor is $  1\, km $.  The number of neutrino  events  per day $ X_{Re}  $
in  the absence of    oscillations  is then given by :
\be
  X_{Re}= X_{Ch} \,   {\Phi}_{ {\bar{\nu} }_e ,\,Re } \,   (V_{Re} /  V_{Ch}) =  1.17 \, d^{-1}
  \ee
Our computation  predicts ( see eq.(\ref{pnuetonueav})) a counting rate  about 3 times lower.

\vspace{10mm}

We are  grateful   to Georges Waysand for a very instructive conversation about the LSSB laboratory and
 for providing informations about the  Reilhanette   unoccupied cavity. We thank John Iliopoulos for his interest
and encouragement.


\begin{thebibliography}{99}
 \bibitem{sno}SNO Collaboration, Q.R. Ahmad {\it et al.}, Phys. Rev. Lett. {\bf 89}, 011301\,  (2002).
 \bibitem{kamland} Kamland  Collaboration, K. Eguchi    {\it et al.}, Phys. Rev. Lett. {\bf 90},  0211802 \,  (2003).
   \bibitem{buch} W. Buchmuller  arXiv:hep-ph/204288 v2  30 May 2002.
    \bibitem{bilen}S.M.  Bilenky, C. Giunti and W. Grimus      arXiv:hep-ph/9812360 v4 4 Jun 1999.
  \bibitem{fog1} G.L. Fogli, E. Lisi, A. Marrone, A. Palazzo and A.M. Rotunno \\
  arXiv:hep-ph/021227 v2  5 Feb 2003.
 \bibitem{fog2}  G.L. Fogli, E. Lisi , A. Marrone and D. Montanino    \\
  arXiv:hep-ph/0303061 v1   7 Mar 2003.
  \bibitem{bah} J.N.  Bahcall, M.C. Gonzalez-Garcia ,  C. Pe\~na-Garay \\
   arXiv:hep-ph/0212147 v3  5 Feb  2003.
    \bibitem{chooz}  Chooz collaboration,  M. Appolino   {\it et  al.},  \\
    arXiv:hep-ex/0301017 v1   13 Jan      20003.
   \bibitem{sk} SK collaboration,Y. Fukuda    Phys. Rev. Lett. {\bf 81},   1562 \, (1998)
  \bibitem{k2k} K2K Collaboration,  S.H. Ahn  {\it et  al.}, Phys. Rev. Lett. {\bf 90}, 041801 \,(2003).

   \end{thebibliography}
   \end{document}